# *On-surface synthesis and characterization of Teranthene and Hexanthene: Ultrashort graphene nanoribbons with mixed armchair and zigzag edges*

Gabriela Borin Barin[1][*][#], Marco Di Giovannantonio[1][*][+], Thorsten G. Lohr[2][*], Shantanu Mishra[1][++], Amogh Kinikar[1], Mickael L. Perrin[3,4], Jan Overbeck[3], Michel Calame[3,5,6], Xinliang Feng[2,7], Roman Fasel[1,8] and Pascal Ruffieux[1][#]

[1]nanotech@surfaces Laboratory, Empa, Swiss Federal Laboratories for Materials Science and Technology, 8600 Dübendorf, Switzerland

[2]Center for Advancing Electronics Dresden, Department of Chemistry and Food Chemistry, TU Dresden, Dresden 01062, Germany

[3]Transport at Nanoscale Interfaces Laboratory, Empa, Swiss Federal Laboratories for Materials Science and Technology, 8600 Dübendorf, Switzerland

[4]Department of Information Technology and Electrical Engineering, ETH Zurich, 8092 Zurich, Switzerland

[5]Department of Physics, University of Basel, Klingelbergstrasse 82, CH-4056 Basel, Switzerland

[6]Swiss Nanoscience Institute, University of Basel, Klingelbergstrasse 82, CH-4056 Basel, Switzerland

[7]Max Planck Institute of Microstructure Physics, Weinberg 2, 06120 Halle, Germany

[8]Department of Chemistry, Biochemistry and Pharmaceutical Sciences, University of Bern, 3012 Bern, Switzerland

[+]Present address: Istituto di Struttura della Materia – CNR (ISM-CNR), via Fosso del Cavaliere 100, Roma 00133, Italy

[++]Present address: IBM Research – Zurich, Rüschlikon 8803, Switzerland

*These authors contributed equally to this work

#corresponding authors: gabriela.borin-barin@empa.ch , pascal.ruffieux@empa.ch

**Abstract**

Graphene nanoribbons (GNRs) exhibit a broad range of physicochemical properties that critically depend on their width and edge topology. While the chemically stable GNRs with armchair edges (AGNRs) are semiconductors with width-tunable band gap, GNRs with zigzag edges (ZGNRs) host spin-polarized edge states, which renders them interesting for applications in spintronic and quantum technologies. However, these states significantly increase their reactivity. For GNRs fabricated via on-surface synthesis under ultrahigh vacuum conditions on metal substrates, the expected reactivity of zigzag edges is a serious concern in view of substrate transfer and device integration under ambient conditions, but corresponding investigations are scarce. Using 10-bromo-9,9':10',9''-teranthracene as a precursor, we have thus synthesized hexanthene (HA) and teranthene (TA) as model compounds for ultrashort GNRs with mixed armchair and zigzag edges, characterized their chemical and electronic structure by means of scanning probe methods, and studied their chemical reactivity upon air exposure by Raman spectroscopy. We present a detailed identification of molecular orbitals and vibrational modes, assign their origin to armchair or zigzag edges, and discuss the chemical reactivity of these edges based on characteristic Raman spectral features.

**Introduction**

The interest in graphene nanoribbons (GNRs) has surged since graphene was first isolated, largely due to the potential for fine-tuning their physicochemical properties by manipulating their size and shape[1,2]. The tailored fabrication of these one-dimensional systems, employing atomically precise control over their width and edge structure, enables the synthesis of GNRs with variable electronic band gaps[3–7], specific topological quantum phases[8,9] or even spin-polarized edge states[10] in a single material class. The required atomic precision for such extended structures is achieved in a bottom-up approach based on the metal surface-catalyzed covalent assembly of carefully designed molecular precursors, which is nowadays called 'on-surface synthesis'. Not only does this approach enable the deterministic synthesis of GNRs (and other nanographenes) with the desired electronic and magnetic properties [8,9,11–13], it also opens up a great potential for their use in nanoelectronics[14–16] and spintronics[17,18].

In order to fully leverage the GNRs properties into device platforms, a transfer step is necessary to bring them from their growth substrate to the technologically relevant ones[15,19]. Next, a careful *out-of-vacuum* characterization is required to ensure that the GNRs' intrinsic properties remain preserved throughout the transfer and device integration process. Significant progress has been made in the *ex situ* characterization, mostly by spectroscopic methods[20–26], and in substrate transfer procedures of armchair-GNRs (AGNRs). Their robustness and stability under ambient conditions[19,27] has allowed the realization of several types of AGNR field effect transistor (FET) devices[28–34].

Moreover, GNRs (and other nanographenes) containing zigzag edges host spin-polarized edge states[10,35–38] which render them interesting for applications in spintronics[39] and quantum computation[40,41]. However, proper *ex situ* characterization and device integration are extremely challenging since zigzag edges are prone to oxidation[13,42–44], which so far has hampered any experimental investigation of their transport characteristics.

Here we designed and synthesized two ultrashort GNR segments, namely teranthene (**TA**) and hexanthene (**HA**). With a comparable number of armchair and zigzag edge sites, TA and HA serve as model systems for the investigation of structure dependent electronic properties and chemical reactivity, in particular compared to long GNRs. **TA** and **HA** are synthesized from the precursor molecule 10-bromo-9,9':10',9''-teranthracene (**1**), by using dehalogenative aryl-aryl coupling and cyclodehydrogenation of **1** on a gold surface[3,45]. We employ low-temperature scanning tunneling microscopy (STM) and non-contact atomic force microscopy (nc-AFM) under ultra-high vacuum (UHV) for the *in situ* investigation of structural and electronic properties of **TA** and **HA** at sub-molecular resolution. Moreover, we utilize vacuum Raman spectroscopy to probe the vibrational signatures of these pristine, atomically precise GNR segments. Characterizing the Raman modes of both **TA** and **HA**, before and after their ambient exposure, allows us to correlate changes in Raman spectroscopy signatures with the chemical reactivity associated with zigzag edges. Our results confirm that Raman spectroscopy can serve as a reliable method to detect the chemical degradation of highly reactive GNRs.

**Results and Discussion**

Ultrashort GNR segments, **TA** and **HA**, were achieved via the precursor **1** (10-bromo-9,9':10',9''-teranthracene) Fig. 1, (the details for the solution synthesis of **1** are in the supporting information, Fig. S1-S3). **1** was sublimated onto an Au(111) surface under UHV conditions and thermally activated to induce surface-assisted dehalogenation and subsequent C-C coupling. 80% of the dehalogenated precursor molecules underwent dimerization at 210° C, while 20% of the surface-stabilized radicals were passivated by hydrogen after dehalogenation of the precursor, thus preventing dimer formation (see Fig. S4 for STM images). Subsequently, intramolecular cyclodehydrogenation of the dimers and passivated monomers was induced by annealing the sample to 290 °C (Fig. 1b-d), which led to a majority of fully planar species coexisting with few partially cyclodehydrogenated molecules (featuring bright protrusions in Fig. 1c). To identify the resulting products, we conducted bond-resolved imaging using nc-AFM with a CO-functionalized tip[46] (Fig. 1d and Fig.S5), which conclusively proves the successful formation of **TA** and **HA**.

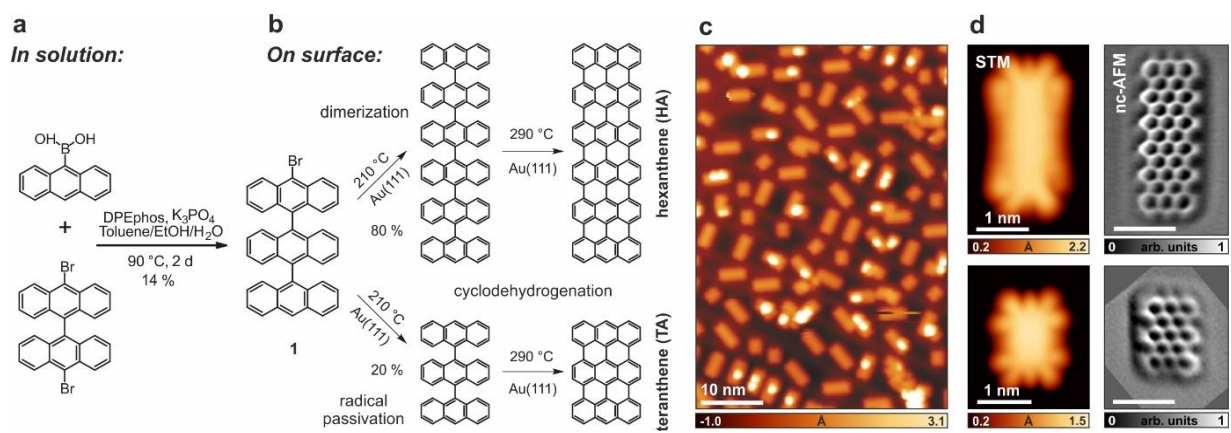

**Figure 1**. a) Solution-synthetic route toward precursor **1**, b) Schematic description of the on-surface synthesis of **TA** and **HA** via dehalogenation and aryl-aryl covalent coupling (radical passivation in the monomer case) followed by cyclodehydrogenation of **1**. c) Large-scale STM image of the surface after deposition of **1** on Au(111) at RT and subsequent annealing to 290 ºC. d) High-resolution STM (left panels) and Laplace-filtered nc-AFM frequency shift images (right panels) of **TA** (bottom) and **HA** (top) obtained using a CO-functionalized tip. Scanning parameters: c: $V = -1.00$ V, $I = 20$ pA, d (STM, top): $V = -20$ mV, $I = 100$ pA, d (STM, bottom): $V = -20$ mV, $I = 200$ pA. See Figure S5 for the corresponding unfiltered nc-AFM images.

The electronic structure of **TA** and **HA** was probed using scanning tunneling spectroscopy (STS). Differential conductance ($dI/dV$) spectra acquired on **TA** reveal four distinct peaks at -1.65 V, -0.10 V, 0.20 V and 1.00 V (Fig. 2a–d). In order to assign these states to specific molecular orbitals, we record spatially resolved $dI/dV$ maps over **TA** at the respective bias voltages and compared them to gas-phase tight-binding local density of state (TB-LDOS) maps of the molecular orbitals of **TA** (Fig. S6 and S7). As shown in Fig. 2d, the $dI/dV$ maps at -0.10 V, 0.20 V and 1.00 V concur with the TB-LDOS maps of the HOMO, LUMO and LUMO+1 of **TA**, thus allowing for an unambiguous assignment of the frontier molecular orbitals. The $dI/dV$ map at -1.65 V, while containing features of both the HOMO–1 and HOMO–2 of **TA** (Fig. S8), cannot be uniquely assigned to a single orbital resonance. Our TB simulations indicate that this resonance likely results from an intermixing of the HOMO–1 and HOMO–2, as has also been previously observed in related molecular systems[47] (Fig. 2d and S8). We also explored the electronic structure of **HA** via STS, where five resonances at -1.10 V, 0.10 V, 0.44 V, 1.25 V and 1.65V are clearly resolved in $dI/dV$ spectroscopy (Fig. 2e,f). Based on the reasonable correspondence between the $dI/dV$ maps at the above biases and the TB-LDOS maps of **HA**, we assign the resonances at –1.10 V, 0.10 V, 0.44 V, 1.25 V and 1.65 V to the gas phase HOMO–1, HOMO, LUMO, LUMO+1 and LUMO+2 of **HA**, respectively (Fig. 2g; also see Fig. S6 and S9). The experimentally measured resonance at 0.10 V, which matches the gas phase HOMO of **HA**, lies above the Fermi energy. This is in contrast with what was observed for **TA** and corresponds to the situation where electron transfer from **HA** to the Au(111) surface leads to the emptying of the HOMO, thus leading to its upshift above the Fermi energy. Our present results are in agreement with previous works, where hole-doping of 7-AGNR segments on Au(111) has been observed by both local scanning probe[48,49] and ensemble electron spectroscopy and photoemission measurements[50]. However, while the reported STS data of extended 7-AGNRs on Au(111) show a band gap of 2.3 eV[51], we observe a significantly narrower gap in the case of **TA** and **HA** on Au(111) (0.30 and 0.34 eV, respectively). This gap reduction is likely due to the hybridization of the localized end states when the zigzag termini of the ribbon become close to one another. Conductance $dI/dV$ spectra of a suspended 7-AGNR revealed that such end states extend up to ~ 1.1 nm from the ribbon terminus[52], which supports their overlap in both **TA** and **HA,** where the total molecular length amounts to 1.4 and 2.8 nm (STM-measured), respectively . This effect modifies the electronic properties of our model compounds with respect to those of longer 7-AGNRs, similarly to what has been reported for the distinct transport properties of finite vs long 5-AGNRs[15,31].

A pertinent question that arises with respect to the electronic structure of **TA** and **HA** is their open- (magnetic) or closed-shell (non-magnetic) ground state on Au(111). Both **TA** and **HA** are members of the family of *peri*–condensed anthracenes denoted as anthenes. All anthenes may be chemically drawn as resonance hybrids of closed-shell Kekulé and open-shell biradical forms, with the later form being driven by the gain in the number of aromatic sextets. The smallest member of this family, bisanthene ($C_{28}H_{14}$), is considered a closed-shell system. However, X-ray crystallographic studies of the larger homologues, namely **TA**[35] and quarteranthene ($C_{56}H_{18}$)[53], showed notable deviation of their geometrical structures from those of closed-shell forms. A more direct route to access the electronic structure of open-shell systems at the single-molecule scale is through STS, where open-shell molecules generally exhibit two distinguishing features. First, the open-shell molecules host singly occupied molecular orbitals (SOMOs), which appear similar in both the occupied and unoccupied regimes due to electron tunneling to/from the same orbitals[54,55]. In contrast, the frontier states of closed-shell molecules consist of the hybridized HOMO and LUMO, which present different shapes and symmetries in *dI/dV* mapping. Second, the presence of unpaired electrons in open-shell molecules additionally manifests as characteristic low-energy spectral features in STS such as Kondo resonances or spin excitations[55,56]. Based on our STS measurements on both **TA** and **HA**, we find no indication of SOMOs and/or magnetic excitations in both the species, and we thus ascribe their electronic ground states to be closed-shell on Au(111).

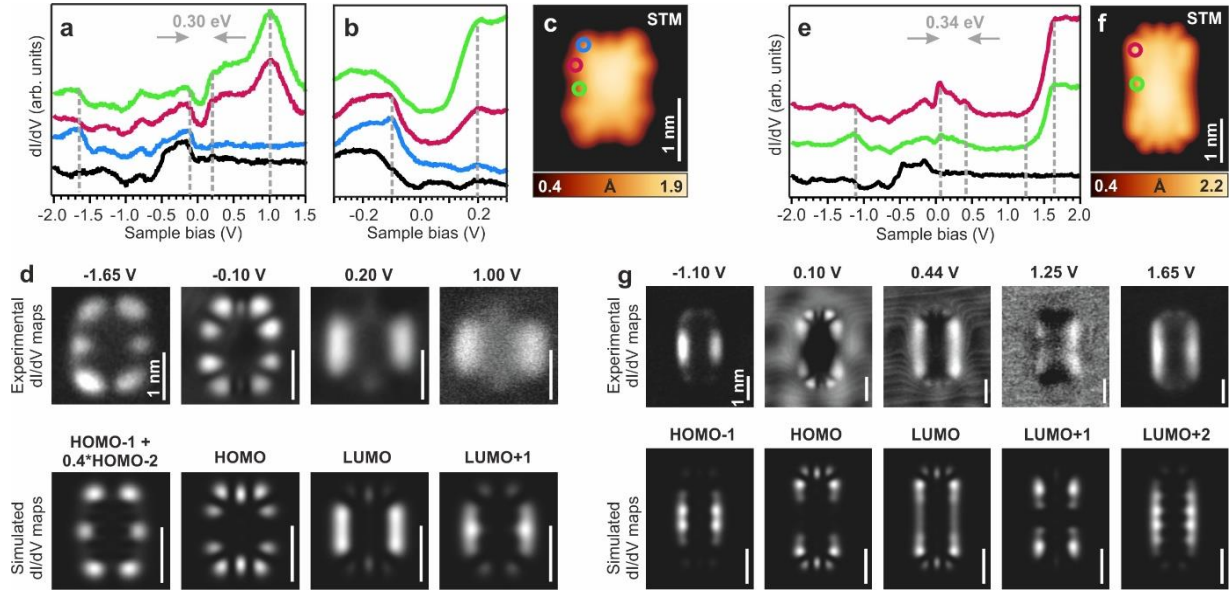

**Figure 2**. Electronic structure characterization of **TA** and **HA**. a) *dI/dV* spectra acquired on **TA** and on the bare substrate (black curve). b) High-resolution *dI/dV* spectroscopy acquired on **TA** in the vicinity of the Fermi level, showing the HOMO and LUMO resonances. c) STM image of **TA** with colored circles indicating the positions where the corresponding *dI/dV* spectra in (a) and (b) were acquired. Scanning parameters: $V$ = –0.10 V, $I$ = 300 pA. d) Upper panels: constant-current *dI/dV* maps recorded at the indicated voltages, corresponding to the main features observed in (a) and (b). Bottom panels: Gas phase TB-LDOS maps of intermixed HOMO-2/HOMO-1, HOMO, LUMO and LUMO+1 of **TA**. e) *dI/dV* spectra acquired on **HA** and on the bare substrate (black curve). f) STM image of **HA** with the colored circles indicating the positions where the corresponding d*I*/d*V* spectra in (e) were acquired. Scanning parameters: $V$ = 0.05 V, $I$ = 100 pA. g) Upper panels: constant-current *dI/dV* maps recorded at the indicated voltages, corresponding to the main spectral features observed in e). Bottom panels: Gas phase TB-LDOS maps of the HOMO-1, HOMO, LUMO, LUMO+1 and LUMO+2 of **HA**.

Surface-averaging techniques like Raman spectroscopy offer an overview of the sample quality at large scales[19] while still being sensitive to the atomic structure of the investigated materials. Raman is a particularly powerful technique, providing information on defects[57], chirality[58], width[21] and length[20] of sp$^2$ hybridized carbon nanostructures. Here, we use Raman spectroscopy to investigate the fingerprint vibrations of **TA** and **HA**. We develop a portable UHV chamber that can be mounted onto the UHV systems in which the samples were prepared and then transferred onto the scanning stage of our Raman microscope. Details of the construction and characterization of the UHV Raman suitcase will be presented elsewhere. This UHV Raman suitcase allows us to characterize the Raman spectroscopy of the pristine GNRs under ultra-high vacuum conditions, and thus allows us to characterize the influence of the environment (UHV vs ambient conditions) on their atomic structure.

The Raman spectrum of a sample with high surface coverage of **TA** and **HA** (Fig. 3a) measured in ultra-high vacuum (~ $5 \cdot 10^{-8}$ mbar) is shown in Fig. 3b. For both, **TA** and **HA**, the high-frequency range of the Raman spectrum is dominated by a peak at 1600 cm$^{-1}$ that is related to the nanographenes' sp$^2$ lattice (G mode). Additionally, we observe the D mode at 1346 cm$^{-1}$ and C-H bending modes at 1220, 1252 and 1274 cm$^{-1}$, proving hydrogen passivation of **TA** and **HA** edges[59–61].

The inset in Fig. 3b shows a zoom-in of the D and C-H region, where a peak at 1142 cm$^{-1}$ and a peak at 1170 cm$^{-1}$ are distinguished. These peaks corresponds to the C-H bending mode of the zigzag edges[62], of **TA** and **HA**, respectively, as indicated by our normal mode analysis (Fig. S10). The C-H bending mode of the zigzag edges of the ribbon termini was so far not observed experimentally due to the low ratio between zigzag (ZZ) and armchair (AC) edge segments of the structures investigated to date. For **TA** and **HA**, we have a much higher proportion of zigzag edges due to their short length: 3 zigzag edge segments per 3 armchair edge segments (ZZ/AC=1) for **TA** and 3 zigzag edge segments per 6 armchair edge segments (ZZ/AC=½) for **HA** (see figure S10). Raman simulations for both **TA** and **HA** show this mode at 1146 cm$^{-1}$ and 1158 cm$^{-1}$, respectively, with higher normalized (to $I_G$) intensity for the teranthene structure (0.11 vs 0.05), in line with the higher ZZ/AC ratio for **TA**.

The low frequency region shows the fingerprint modes of both width and length of the species on the surface. The fundamental transverse acoustic mode for GNRs, called radial breathing-like mode (RBLM), is specific for each GNR width[21,63]. Both **TA** and **HA** may be structurally seen as short 7-AGNRs, and here we observe that their RBLM is in the same region than the one of long 7-AGNRs[64], around 394 cm$^{-1}$, but with a different peak shape. A similar behavior was observed for short 5-AGNRs (1-2.5 nm) where the RBLM showed side-peaks due to the splitting of the RBLM into normal modes with diagonal atomic displacement[20].

Recently we showed that in a sample containing AGNRs with broad length distribution, probing the longitudinal compressive mode (LCM) of AGNRs with Raman spectroscopy enables us to obtain surface-averaged length information on the growth substrate and upon transfer to technological relevant substrates[20]. Here, in virtue of the controlled selectivity of our precursor design, the short length of both **TA** and **HA** allows us to observe two characteristic length-dependent modes, one at 127 cm$^{-1}$ for **HA** and another at 246 cm$^{-1}$ for **TA**, in agreement with their DFT simulated spectra (Fig. 3c). In line with the STM images, which shows 80% of **HA** on the surface, the LCM of this species has higher intensity relative to the **TA** (0.1 vs 0.02 – intensity normalized to $I_G$).

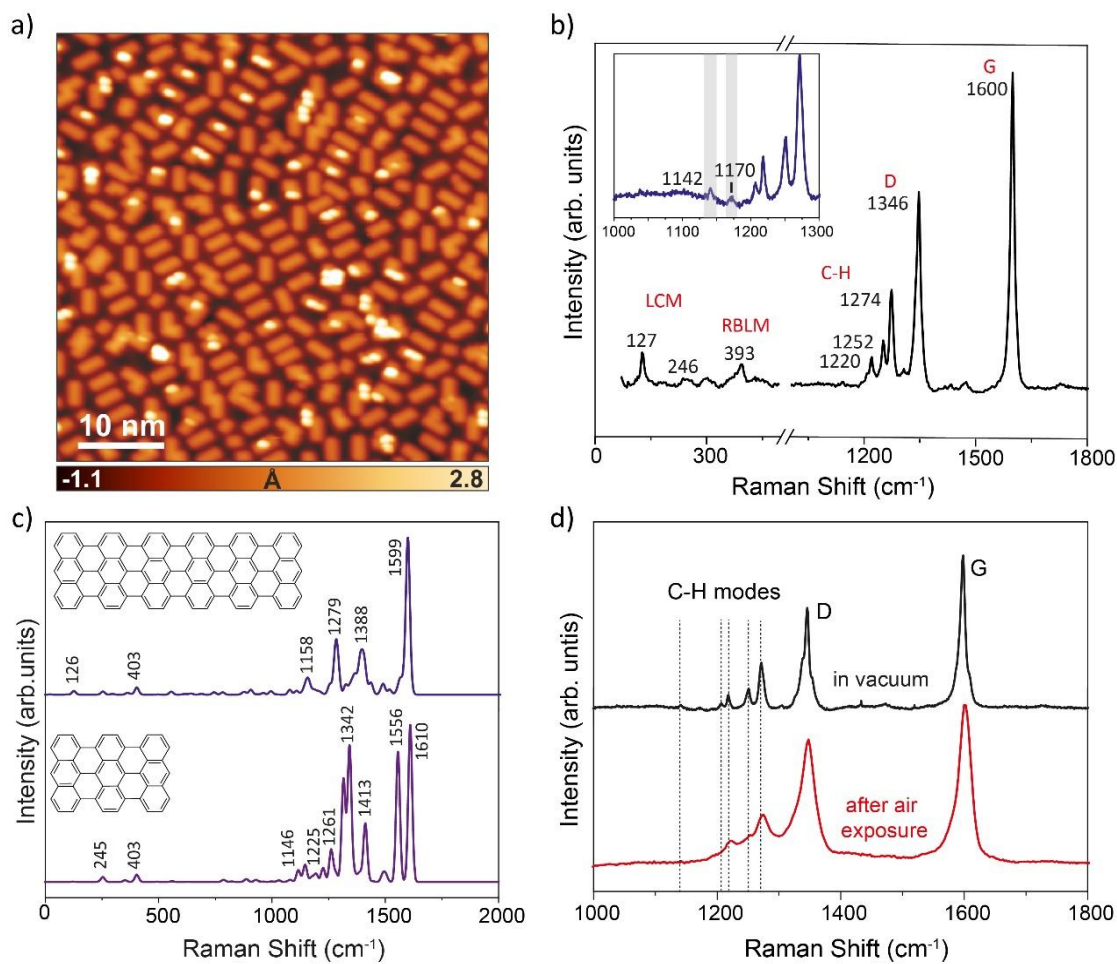

**Figure 3**. a) STM topography of a high coverage sample of **1** on Au/mica, annealed to 290 ºC. Scanning parameters: $V = -1$ V, $I = 20$ pA), b) Raman spectroscopy of the same sample as shown in a), measured in ultra-high-vacuum with a 532 nm laser. Inset shows a high resolution spectrum in the D region highlighting the C-H bending mode of zigzag edges. c) DFT simulated Raman spectra of **TA** and **HA** d) Comparison between Raman spectra acquired on the sample kept in UHV (black) and after exposure to air (red).

**TA** and **HA** have a larger ZZ to AC edge ratio in comparison to long 7-AGNRs, and thus have easily detectable zigzag edge vibrational modes at 1142 cm$^{-1}$ (from **TA**) and 1170 cm$^{-1}$ (from **HA**) in Raman spectroscopy (Fig. 3b). Therefore, they serve as excellent model systems to investigate the chemical reactivity of nanographenes/GNRs upon exposure to air.

Figure 3d shows the high frequency Raman profiles of the edge-related modes (D and CH bending) and G mode measured in vacuum, with (in red) or without (in black) an air-exposure step in between. The Raman profiles reveal broadening and disappearance of the edge-related modes for the sample briefly (~5 minutes) exposed to air. Particularly we observe that the CH-bending modes assigned to the zigzag edges are not resolved anymore after air exposure along with broadening of all other CH-bending related modes as shown by their increased full-width at half maximum: mode at ~1220 cm$^{-1}$ from FWHM = 5 cm$^{-1}$ to 19 cm$^{-1}$, mode at ~1274 cm$^{-1}$ from FWHM = 9 cm$^{-1}$ to 43 cm$^{-1}$, both D modes from FWHM = 9 cm$^{-1}$ to 27cm$^{-1}$, and G mode from FWHM=10 cm$^{-1}$ to 23 cm$^{-1}$. This behavior is not observed for long 7-AGNRs where peak widths remains the same upon air exposure (Refs. [19,64] and figure S11).

The FWHM of Raman peaks is a useful tool to characterize sample crystallinity or purity. Recent work by Lawrence *et al*. demonstrated that exposing chiral (3,1)-GNRs to air results in their oxidation[43]. Moreover, this process generates several oxidized species, undermining the atomic precision achieved through on-surface synthesis. Considering that the armchair edges of the 7-AGNR are chemically unreactive, any oxidation is limited to their zigzag termini[13,42]. The signal from these zigzag termini are negligible compared to the armchair edges because the 7-AGNRs have an average length exceeding 15-20 nm. However, by choosing to study the ultrashort 7-AGNR segments, we have amplified the signal of the zigzag edges with respect to the armchair edges. This allows us to detect their presence when the Raman is measured in UHV and their absence upon air exposure.

Furthermore, due to the small size of these segments, any chemical change such as ketone or carboxyl substitution, would have a significant effect on all vibrational modes in **TA** and **HA**. We expect a variety of such uncontrolled chemical changes to occur upon air exposure, yielding several different oxidized species, each with their own peak shifts (see Figure S12 for a simulated spectra of both TA and HA with OH-terminated zigzag edges, showing new peak shifts and broadening). Cumulatively, this broadens all the Raman peaks when compared with the pristine samples measured in UHV.

Our findings reaffirm the chemical reactivity of the zigzag edges. We also establish that Raman spectroscopy is an extremely efficient technique for characterizing the chemical degradation occurring when these edges oxidize. Consequently, Raman spectroscopy serves as an effective tool for verifying the chemical integrity of systems at different stages of device integration. Our results underscore the inherent challenges in transferring GNRs with zigzag edges, emphasizing the need for novel fabrication and encapsulation methods to preserve these edges and the related spin-polarized edge states for their potential integration into spintronic platforms.

**Conclusions**

In conclusion, we successfully synthesized the ultrashort 7-AGNR model compounds teranthene (**TA**) and hexanthene (**HA**) from 10-bromo-9,9':10',9''-teranthracene precursor on

Au(111), and unveiled their atomically-resolved chemical structures by STM and nc-AFM. Detailed electronic characterization and tight-binding calculations revealed electronic gaps of 0.34 eV and 0.30 eV between frontier orbitals of **HA** and **TA** on Au(111), respectively, with the former being positively charged on this substrate. Our precursor design allowed assessing Raman fingerprints of the length-dependent vibrational modes of **TA** and **HA** at 245 and 127 cm$^{-1}$, respectively, which would be impossible with standard di-halogenated precursors that produce polymers and GNRs with broad length distributions. In virtue of the high ZZ to AC edge ratio of the nanostructures synthesized herein, we also identified for the first time CH-bending modes of the zigzag-edges at 1142 cm$^{-1}$ for **TA** and 1170 cm$^{-1}$ for **HA.** Finally, we probed the reactivity of **TA** and **HA** by measuring Raman spectra of samples kept in ultra-high vacuum and exposed to air. Our results showed broadening and suppression of the AC and ZZ edge-related modes, respectively, which we associate to oxidation of the highly reactive zigzag edges.

## Methods

### *STM/STS and nc-AFM*

The on-surface synthesis experiments were performed under ultrahigh vacuum (UHV) conditions with base pressure below $2\times10^{-10}$ mbar. The Au(111) single crystal (MaTeck GmbH) was cleaned by several cycles of Ar$^+$ sputtering (1 keV) and annealing (470 °C) until a clean surface with monoatomic terrace steps is achieved. Deposition of the precursor molecule was performed by thermal evaporation from a 6-fold organic evaporator (Mantis GmbH). A commercial low-temperature STM (Scienta Omicron) was used for sample characterization, operated at 5 K in constant-current mode using an etched tungsten tip. Bias voltages are given with respect to the sample. Constant-height d$I$/d$V$ spectra and maps were obtained with a lock-in amplifier (f = 610 Hz). Nc-AFM measurements were performed at 5 K with a tungsten tip placed on a QPlus tuning fork sensor[65]. The tip was functionalized with a single CO molecule at the tip apex picked up from the previously CO-dosed surface[46,66]. The sensor was driven at its resonance frequency (27000 Hz) with a constant amplitude of 70 pm. The frequency shift from resonance of the tuning fork was recorded in constant-height mode using Omicron Matrix electronics and HF2Li PLL by Zurich Instruments.

### *Raman Spectroscopy*

After UHV preparation and characterization, the samples were transferred into a vacuum suitcase via the fast-entry lock of the UHV apparatus. The maximum pressure experienced by the samples during transfer and Raman characterization was in the order of 10$^{-8}$ mbar.

Raman spectra were acquired with a WITec Alpha 300 R confocal Raman microscope in backscattering geometry. Measurements were performed with 532 nm excitation and 150g/mm grating. For maximum signal intensity spectra are recorded with a Zeiss 50x LD objective, NA=0.55, through an uncoated fused silica window of only 0.2mm thickness covering a hole of just 7mm diameter.

*Raman simulations*

The Raman spectra were calculated with the ORCA 4.2 DFT code[67] using the GGA PBE exchange-correlation functional, and the def2-SVP basis set in combination with the RI-J approximation for Coulomb integrals. The frequencies were calculated numerically. All atomic position are obtained from a geometry relaxation performed with the same settings as the Raman calculations.

*Computational details*

**Tight-binding calculations:** The tight-binding calculations of **TA** and **HA** were performed by numerically solving the nearest-neighbor hopping tight-binding Hamiltonian, considering only the carbon $2p_z$ orbitals:

$$\hat{H}_{TB} = -t \sum_{\langle \alpha,\beta \rangle,\sigma} c^\dagger_{\alpha,\sigma} c_{\beta,\sigma} \qquad (1)$$

Here, $c^\dagger_{\alpha,\sigma}$ and $c_{\beta,\sigma}$ denote the spin selective ($\sigma \in \{\uparrow,\downarrow\}$) creation and annihilation operator at neighboring sites $\alpha$ and $\beta$, and $t$ is the nearest-neighbor hopping parameter, with $t = 2.7$ eV used throughout this work.

Orbital electron densities, $\rho$, of the $n^{\text{th}}$-eigenstate with energy $E_n$ have been simulated from the corresponding state vector $a_{n,i,\sigma}$ by:

$$\rho_{n,\sigma}(\vec{r}) = \left| \sum_i a_{n,i,\sigma} \phi_{2p_z}(\vec{r} - \vec{r}_i) \right|^2, \qquad (2)$$

where $i$ denotes the atomic site index, and $\phi_{2p_z}$ denotes the Slater $2p_z$ orbital for carbon.

*Acknowledgments*

This worked was supported by the Swiss National Science Foundation under grant no. 200020_182015, IZLCZ2_170184 and CRSII5_205987, the European Union Horizon 2020 research and innovation program under grant agreement no. 881603 (GrapheneFlagship Core 3), and the Office of Naval Research BRC Program under the grant N00014-18-1-2708. The authors also greatly appreciate the financial support from the Werner Siemens Foundation (CarboQuant). M.C. acknowledges support by the Swiss National Science Foundation under the Sinergia grant no. 189924 (Hydronics) and funding from the EC H2020 FET Open project no. 767187 (QuIET). M.L.P. acknowledges funding from the Swiss National Science Foundation under Spark grant no. 196795 and the Eccellenza Professorial Fellowship no. PCEFP2\textunderscore203663, as well as support by the Swiss State Secretariat for Education, Research and Innovation (SERI) under contract number MB22.00076. Lukas Rotach is gratefully acknowledged for his technical support during the experiments.

For the purpose of Open Access (which is required by our funding agencies), the authors have applied a CC BY public copyright license to any Author Accepted Manuscript version arising from this submission.

*Author Contribution*

Conceptualization: GBB, MDG, TL, PR Data Curation: GBB, MDG, TL, JO, MP, SM ., Formal Analysis: GBB, MDG, TL, SM, PR. Funding Acquisition: RF, MP, PR, XF, MC, Investigation: GBB, MDG, TL. Supervision: RF, PR, MC, XF. Methodology: GBB, MDG, TL, JO, AK. Visualization: GBB, MDG, TL, Writing|original draft: GBB, MDG, TL, SM, PR Writing|review & editing: all

*Conflicts of Interest*

There are not conflicts to declare

# Supporting Information

# *On-surface synthesis and characterization of Teranthene and Hexanthene: Ultrashort graphene nanoribbons with mixed armchair and zigzag edges*

Gabriela Borin Barin[1]*, Marco Di Giovannantonio[1]*+, Thorsten G. Lohr[2]*, Shantanu Mishra[1]++, Amogh Kinikar[1], Mickael L. Perrin[3,4], Jan Overbeck[3], Michel Calame[3,5,6], Xinliang Feng[2,7], Roman Fasel[1,8] and Pascal Ruffieux[1]

[1]nanotech@surfaces Laboratory, Empa, Swiss Federal Laboratories for Materials Science and Technology, 8600 Dübendorf, Switzerland

[2]Center for Advancing Electronics Dresden, Department of Chemistry and Food Chemistry, TU Dresden, Dresden 01062, Germany

[3]Transport at Nanoscale Interfaces Laboratory, Empa, Swiss Federal Laboratories for Materials Science and Technology, 8600 Dübendorf, Switzerland

[4]Department of Information Technology and Electrical Engineering, ETH Zurich, 8092 Zurich, Switzerland

[5]Department of Physics, University of Basel, Klingelbergstrasse 82, CH-4056 Basel, Switzerland

[6]Swiss Nanoscience Institute, University of Basel, Klingelbergstrasse 82, CH-4056 Basel, Switzerland

[7]Max Planck Institute of Microstructure Physics, Weinberg 2, 06120 Halle, Germany

[8]Department of Chemistry, Biochemistry and Pharmaceutical Sciences, University of Bern, 3012 Bern, Switzerland

+Present address: Istituto di Struttura della Materia – CNR (ISM-CNR), via Fosso del Cavaliere 100, Roma 00133, Italy

++Present address: IBM Research – Zurich, Rüschlikon 8803, Switzerland

*These authors contributed equally to this work

**Precursor synthesis and characterization**

Unless otherwise stated, commercially available starting materials, catalysts, reagents, and dry solvents were used without further purification. Reactions were performed using standard vacuum-line and Schlenk techniques. All the starting materials were obtained from Sigma Aldrich, or TCI. Catalysts were purchased from Strem. Column chromatography was performed on silica (SiO$_2$, particle size 0.063−0.200 mm, purchased from VWR). Silica-coated aluminum sheets with a fluorescence indicator (TLC silica gel 60 F$_{254}$, purchased from Merck KGaA) were used for thin layer chromatography.

MALDI-TOF spectra were recorded on a Bruker Autoflex Speed MALDI-TOF MS (Bruker Daltonics, Bremen, Germany). All of the samples were prepared by mixing the analyte and the matrix, 1,8-dihydroxyanthracen-9(10H)-one (dithranol, purchased from Fluka Analytical, purity >98%) in the solid state.

NMR data were recorded on a Bruker AV-II 300 spectrometer operating at 300 MHz for $^1$H and 75 MHz for $^{13}$C with standard Bruker pulse programs 333 K. Chemical shifts ($\delta$) are reported in ppm. Coupling constants ($J$) are reported in Hz. Tetrachloroethane-d$_2$ ($\delta(^1$H$)$ = 5.91 ppm, $\delta(^{13}$C$)$ = 74.2 ppm) was used as solvents. The following abbreviations are used to describe peak patterns as appropriate: s = singlet, d = doublet, t = triplet, q = quartet, and m = multiplet. Tetrachloroethane-d$_2$ (99.9 atom% D) was purchased from Euriso-top.

Melting points were determined on a Büchi Melting Point M-560 in a range of 50−400 °C with a temperature rate of 10 °Cmin$^{-1}$.

10-bromo-9,9':10',9''-teranthracene (**1**)

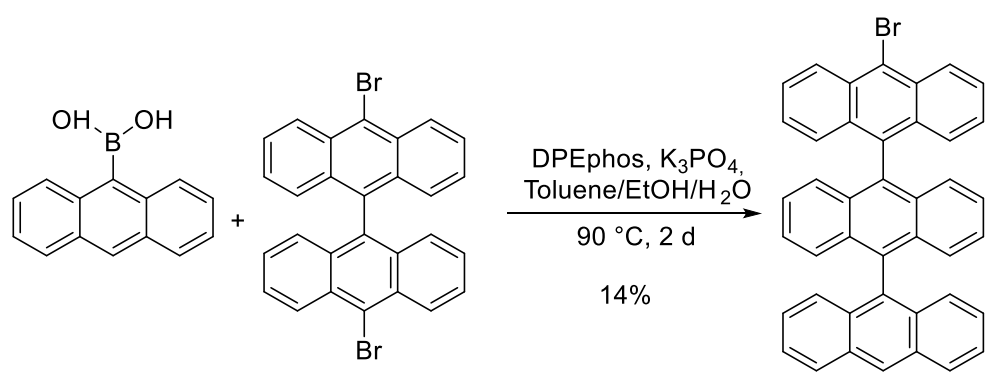

**Figure S1**. 10,10´-dibromo-9,9´-bianthracene (100 mg, 0,195 mmol), 9-anthraceneboronic acid (56.4 mg, 0.254 mmol, 1.3 eq.), tripotassium phosphate (207.2 mg, 0.976 mmol, 5.0 eq.), DPEphos (42 mg, 0.078 mmol, 0.4 eq.) and Pd$_2$(dba)$_3$ (42.1 mg, 0.078 mmol, 0.4 eq.) were added to a 50 ml schlenk flask and evacuated and refilled with argon three times. After adding the degassed solvents (5 ml toluene, 1 ml water and 0.6 ml ethanol), the reaction mixture was heated to 90 °C for 2 days. After quenching with water the product was extracted with DCM, washed with brine and dried over MgSO$_4$. The solvents were removed under vacuum. The crude material was purified by column chromatography (chloroform:isohexane = 1:5 to 1:1). After recrystallization (chloroform/MeOH) the title compound **1** was afforded in a yield of 14% (16 mg).

mp: > 400 °C; R$_f$: 0.30 (DCM:isohexane = 1:3); $^1$H NMR (300 MHz, C$_2$D$_2$Cl$_4$, 333 K) $\delta$ (ppm) = 8.71 (d, $J$ = 6.9 Hz, 3H), 8.16 (d, $J$ = 8.5 Hz, 2H), 7.65 – 7.56 (m, 2H), 7.52 – 7.43 (m, 2H), 7.34 – 7.23 (m, 8H), 7.21 – 7.13 (m, 4H), 7.11 – 7.01 (m, 4H); $^{13}$C NMR (76 MHz, C$_2$D$_2$Cl$_4$, 333K) $\delta$ (ppm) = 132.74, 131.91, 131.66, 128.97, 128.40, 127.79, 127.61, 127.30, 127.06, 126.68, 126.42, 126.00, 125.74; HR-MS (MALDI-TOF) *m/z*: [M$^+$] calcd for C$_{42}$H$_{25}$Br: 608.1130; found: 608.1128; error: -0.33 ppm.

## MALDI-TOF spectra

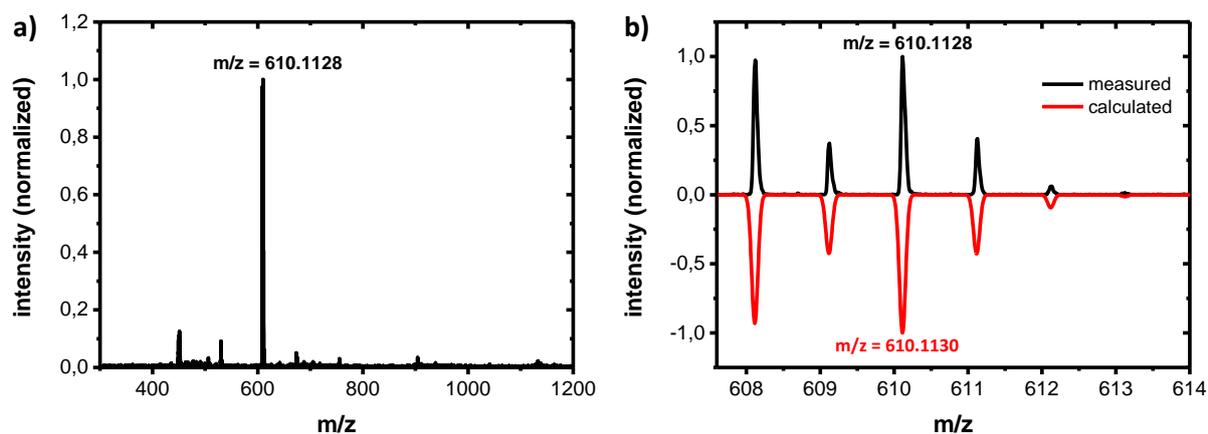

**Figure S2**: a) HR-MALDI-TOF spectrum of **1**; b) HR-MALDI-TOF magnified spectrum of **1** (black line) is in agreement to the expected isotopic distribution pattern (red line).

## NMR spectra

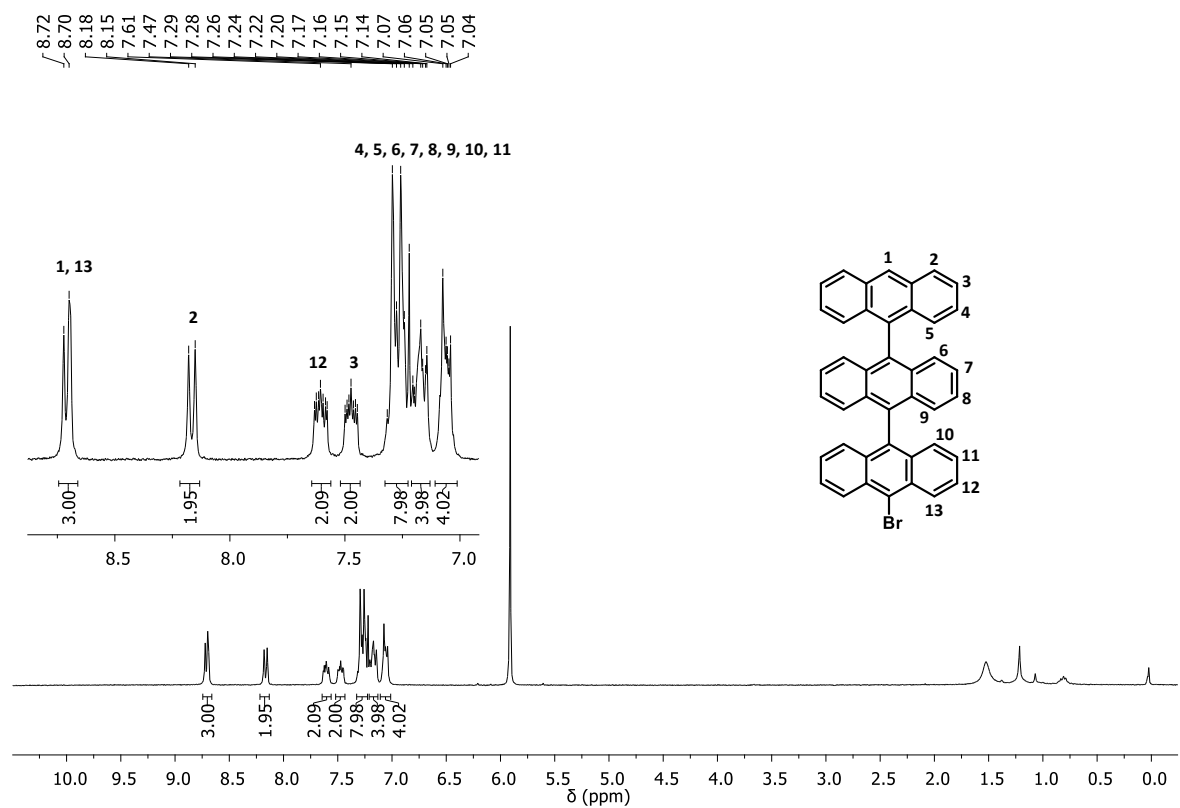

**Figure S3a**: $^1$H-NMR spectrum of **1** dissolved in tetrachloroethane-d$_2$, 300 MHz, 333 K.

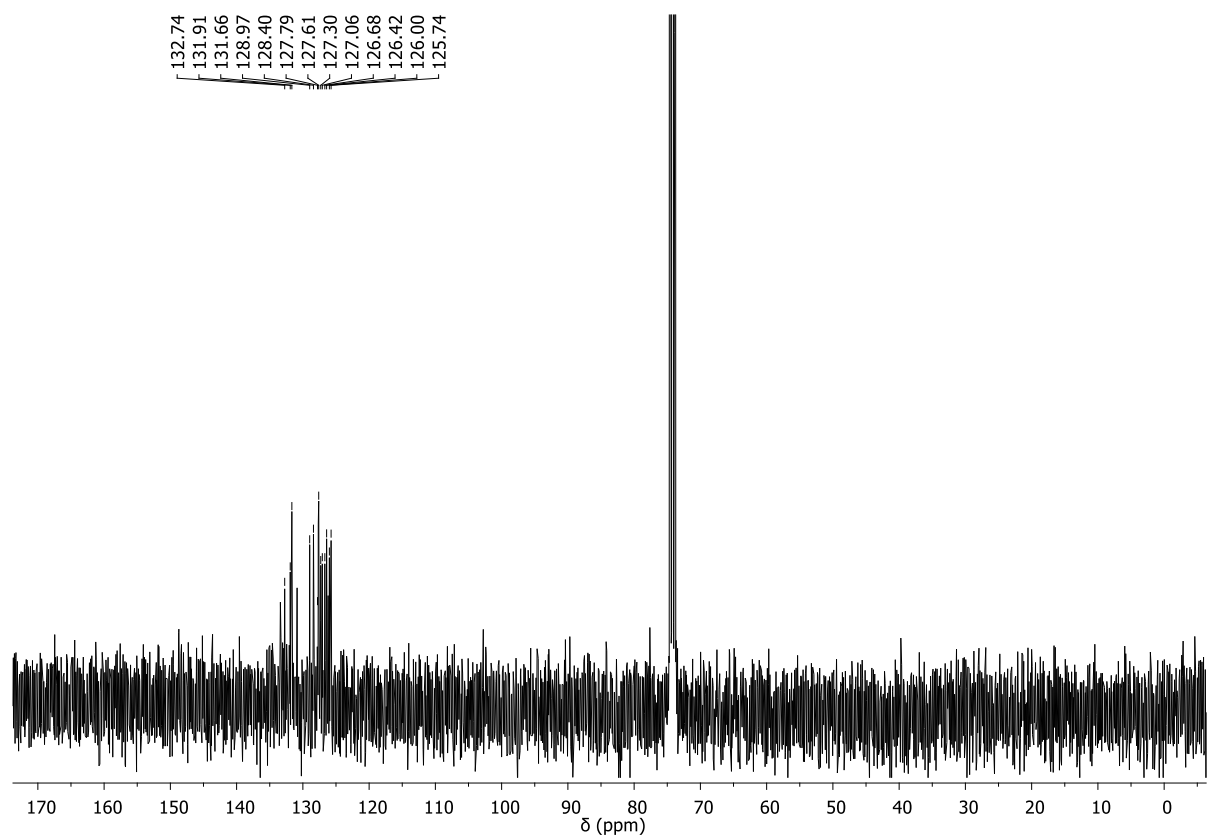

**Figure S3b**: $^{13}$C-NMR spectrum of **1** dissolved in tetrachloroethane-d$_2$, 75 MHz, 333 K.

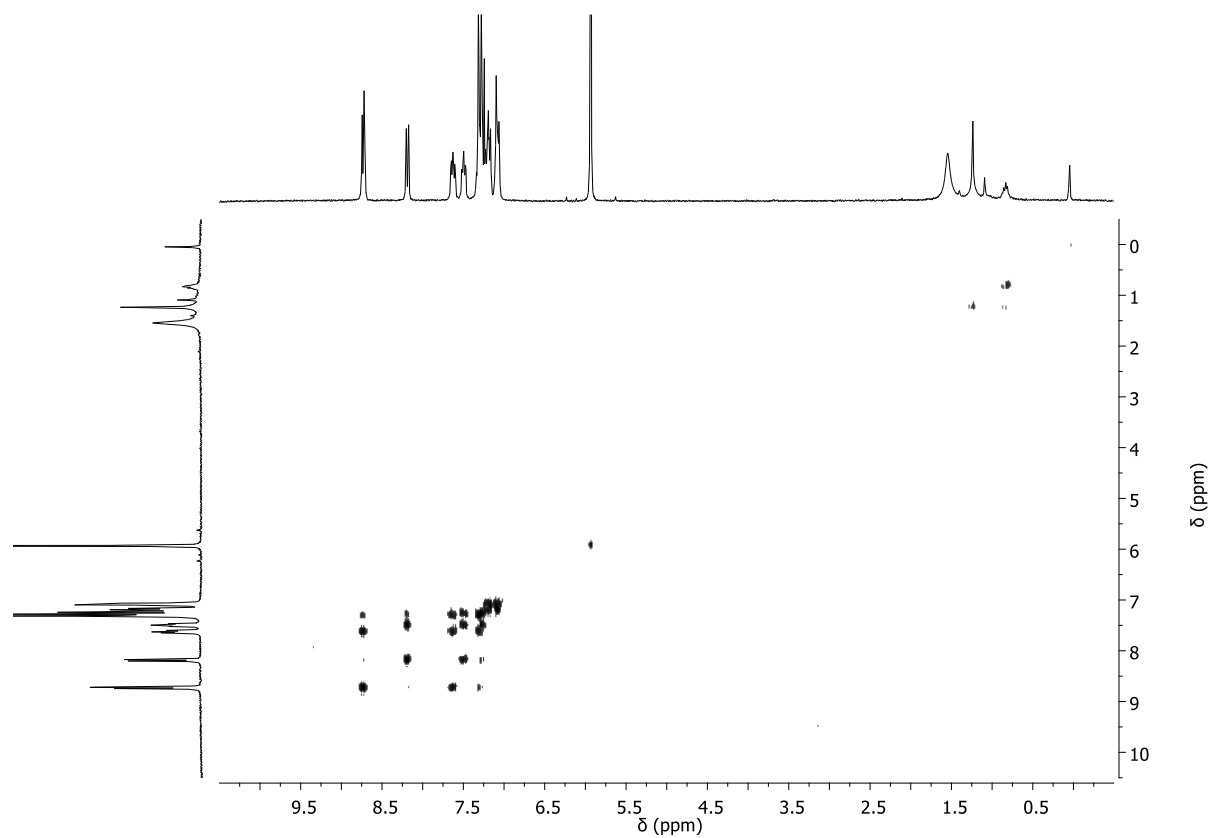

**Figure S3c**: $^{1}$H/$^{1}$H-COSY-NMR spectrum of **1** dissolved in tetrachloroethane-d$_2$, 300 MHz, 333 K.

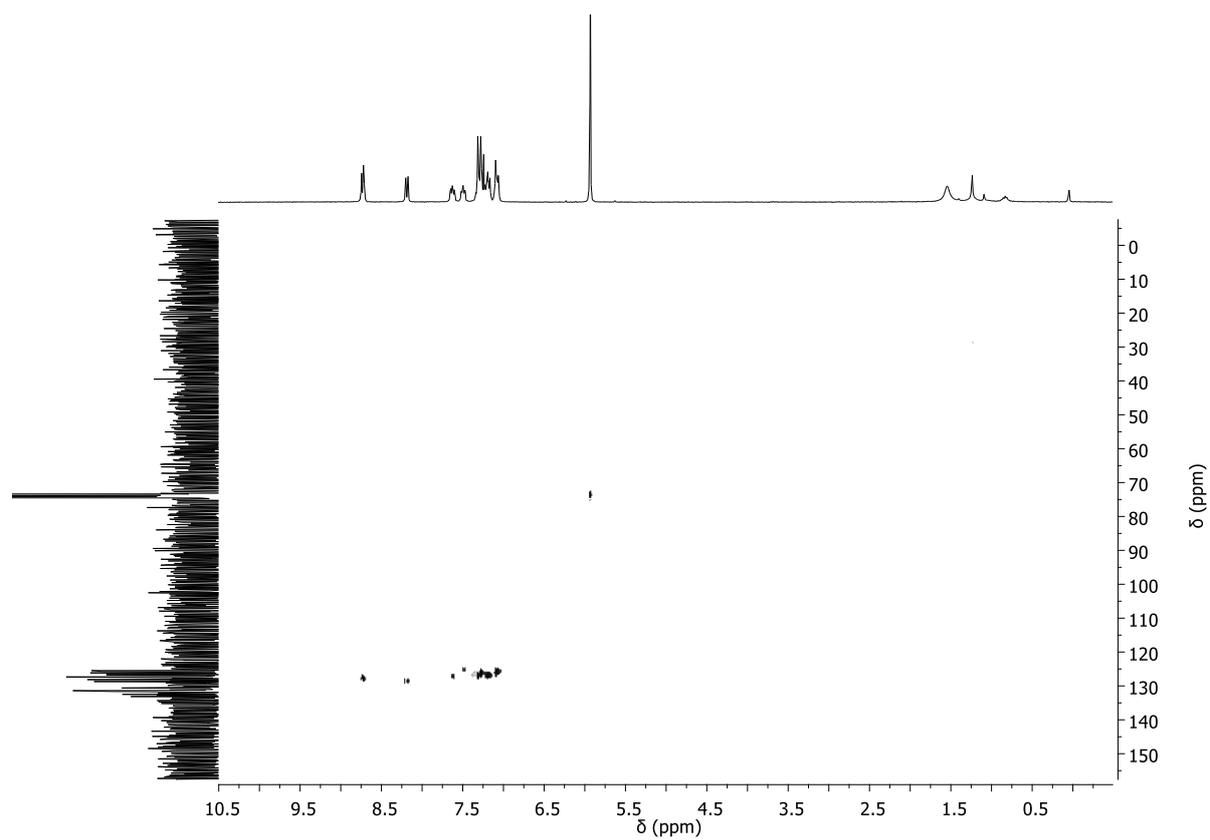

**Figure S3d**: HSQC-NMR spectrum of **1** dissolved in tetrachloroethane-d$_2$, 75 MHz, 333 K.

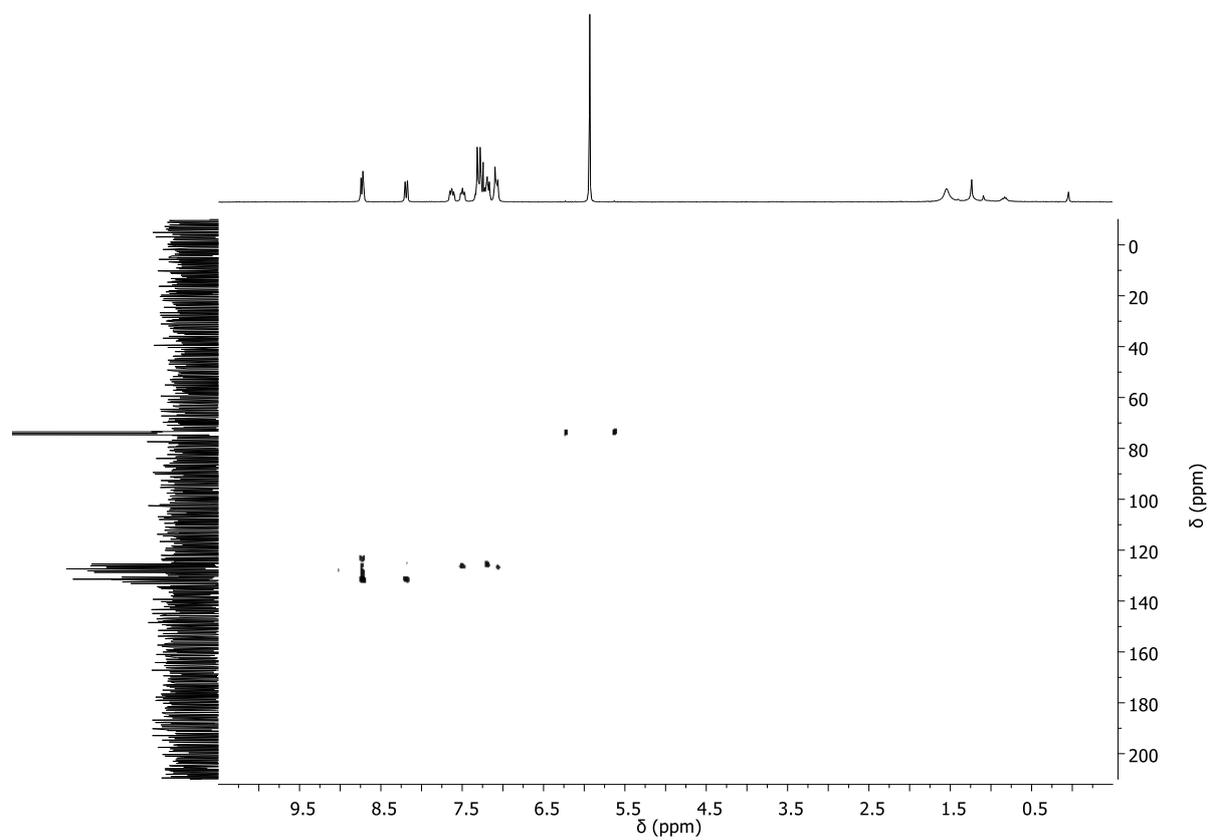

**Figure S3e**: HMBC-NMR spectrum of **1** dissolved in tetrachloroethane-d$_2$, 75 MHz, 333 K.

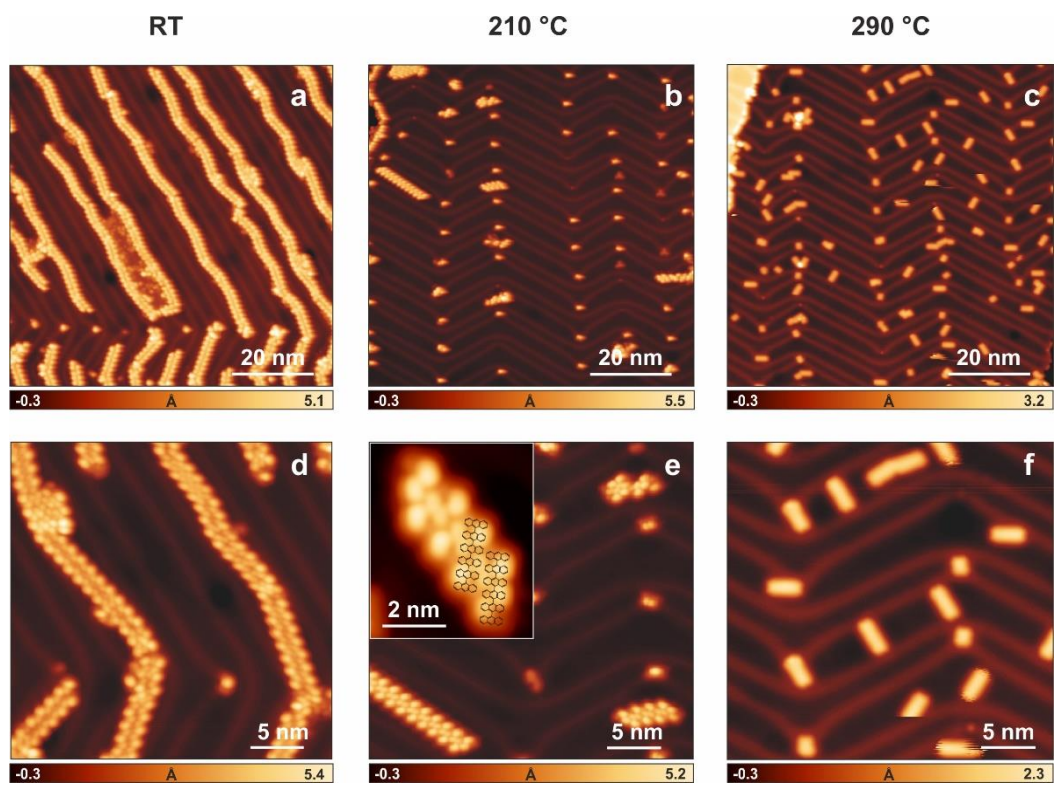

**Figure S4.** STM images of the Au(111) surface after deposition of precursor **1** at room temperature (RT), followed by successive annealing steps at the given temperatures (a-f). Inset (e): magnified STM image with superimposed chemical structures of the dimers after dehalogenative aryl-aryl coupling of **1** on the surface. Scanning parameters: a) $V = -1.0$ V, $I = 30$ pA, b) $V = -1.0$ V, $I = 30$ pA, c) $V = -1.0$ V, $I = 20$ pA, d) $V = -0.5$ V, $I = 30$ pA, e) $V = -0.3$ V, $I = 30$ pA, f) $V = -1$ V, $I = 20$ pA.

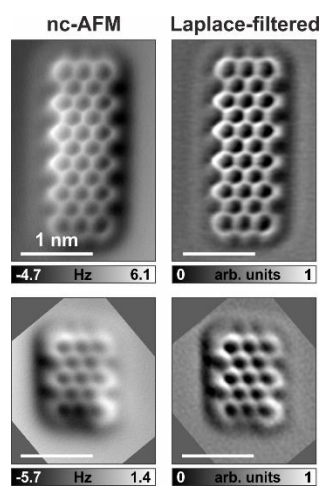

**Figure S5.** Left panels: nc-AFM frequency-shift images of **HA** (upper panel) and **TA** (lower panel). Scanning parameters: (top) tip height offset $\Delta z = 195$ pm above the STM setpoint $V = -5$ mV, $I = 100$ pA, (bottom) $\Delta z = 190$ pm above the STM setpoint $V = -5$ mV, $I = 100$ pA. Right panels: Corresponding Laplace-filtered versions of the left panels (also shown in Fig. 1).

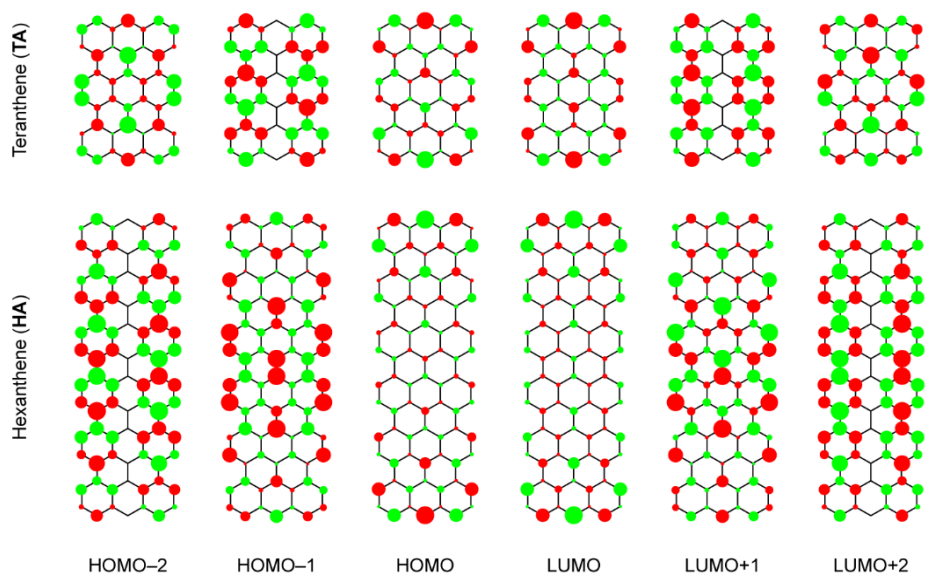

**Figure S6**. TB wave functions of HOMO–2 to LUMO+2 of **TA** (top) and **HA** (bottom). Size of the filled circles denotes amplitude of the wave function, while the two colors denote opposite phases of the wave function.

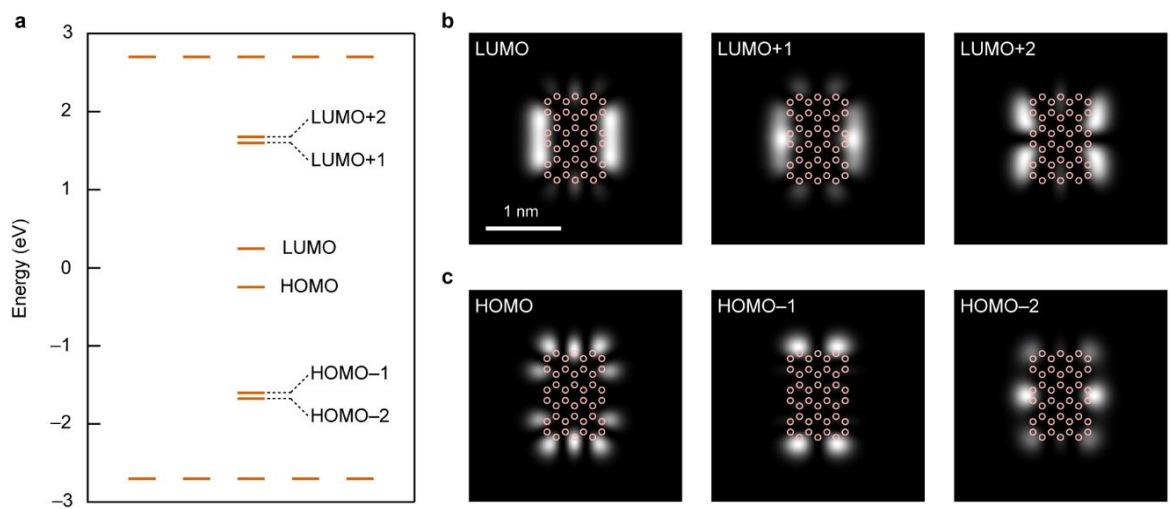

**Figure S7.** TB calculations on **TA**. (a) Gas phase TB energy spectrum of **TA** between ±3 eV. HOMO–2 to LUMO+2 states are labeled. The calculated frontier gap of **TA** is ~0.5 eV at nearest-neighbor TB level of theory. (b, c) TB-LDOS maps of the labeled states in (a) for the unoccupied (b) and occupied (c) channels. The carbon backbone of **TA** (colored circles) is superimposed as a guide to the eye. The maps are calculated at a height of 7 Å above the molecular plane.

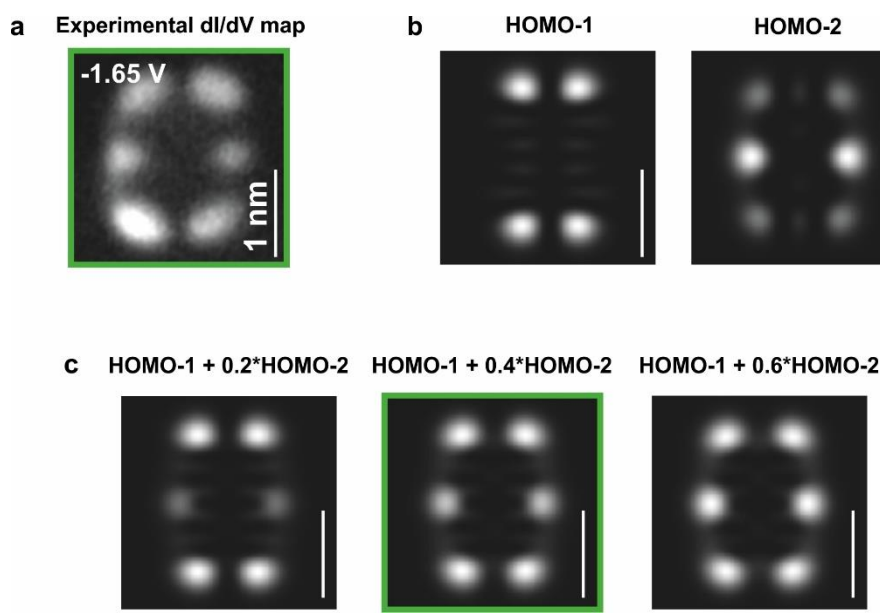

**Figure S8.** Intermixing of HOMO-1 and HOMO-2 states of **TA**. (a) d*I*/d*V* map of the resonance at –1.65 V. (b) TB-LDOS maps of the HOMO–1 and HOMO–2 of **TA**. The d*I*/d*V* map at –1.65 V contains features that are characteristic of both orbitals – that is, strong LDOS features at the termini (characteristic of HOMO–1, and to some extent, also HOMO–2) and slightly weaker LDOS features at the center of the armchair edges (characteristic of HOMO–2 exclusively). As shown in (c), mixing of the HOMO–2 LDOS (with a weight of 0.4) into the HOMO–1 LDOS clearly reproduces our experimental observation. The TB-LDOS maps are calculated at a height of 7 Å above the molecular plane.

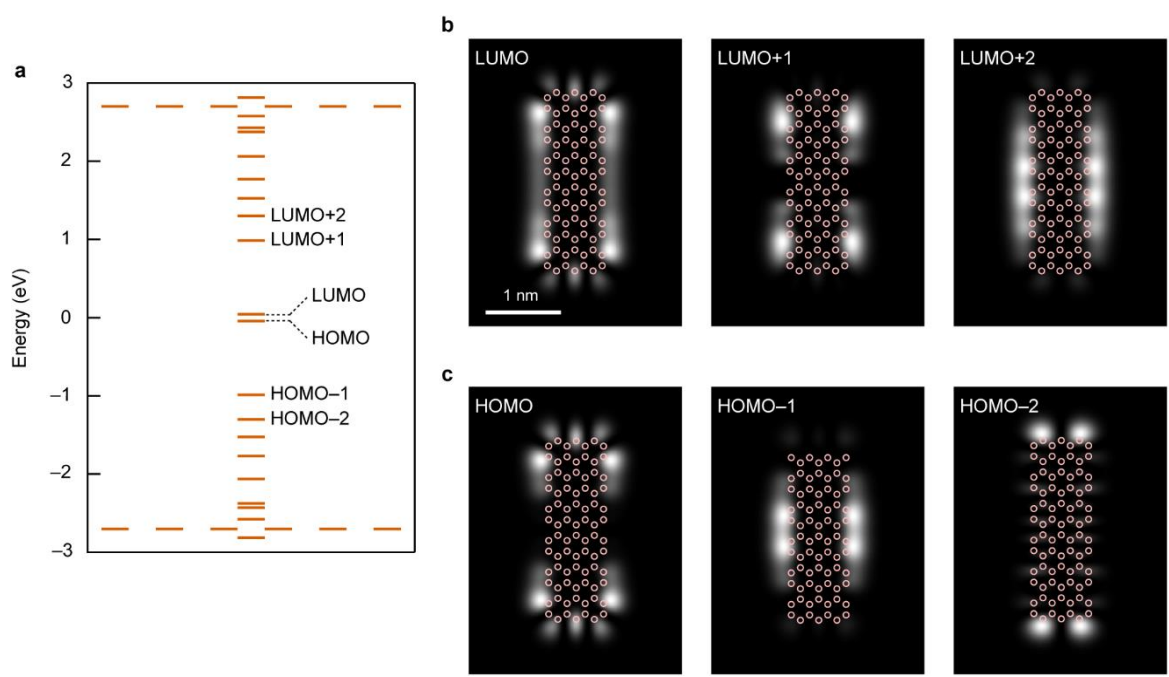

**Figure S9.** TB calculations on **HA**. (a) Gas phase TB energy spectrum of **HA** between ±3 eV. HOMO–2 to LUMO+2 states are labeled. The calculated frontier gap of **HA** is ~90 meV at nearest-neighbor TB level of theory, much less than that of **TA** at the same level of theory (Fig.

S7). (b, c) TB-LDOS maps of the labeled states in (a) for the unoccupied (b) and occupied (c) channels. The carbon backbone of **HA** (colored circles) is superimposed as a guide to the eye. The maps are calculated at a height of 7 Å above the molecular plane.

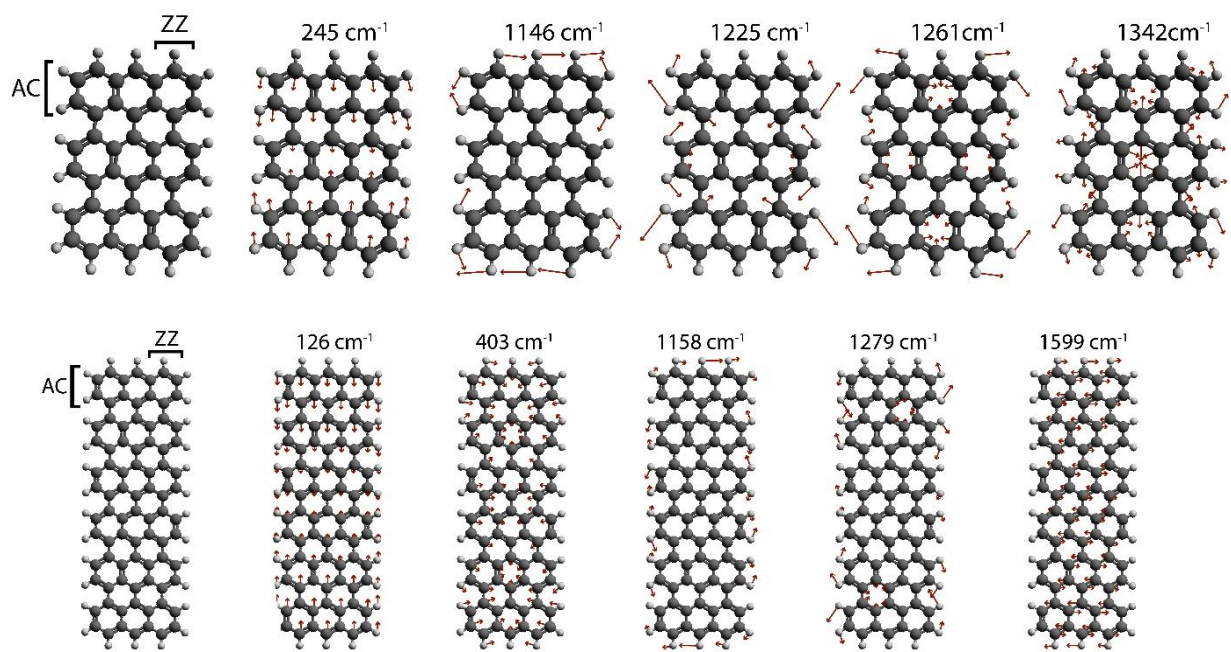

**Figure S10:** Raman normal mode analysis of teranthene (**TA**) (upper panel) and hexanthene (**HA**) (lower panel). Red arrows indicate amplitude and direction of atomic displacements.
- Modes at 245 and 126 cm$^{-1}$: longitudinal compressive mode (LCM)
- Mode at 403 cm$^{-1}$: radial breathing-like mode (RBLM)
- Modes at 1146 and 1158 cm$^{-1}$: CH-bending of zigzag edges
- Modes at 1225, 1261 and 1279 cm$^{-1}$: CH-bending armchair edges
- Mode at 1342 cm$^{-1}$: D mode
- Mode at 1599: G mode

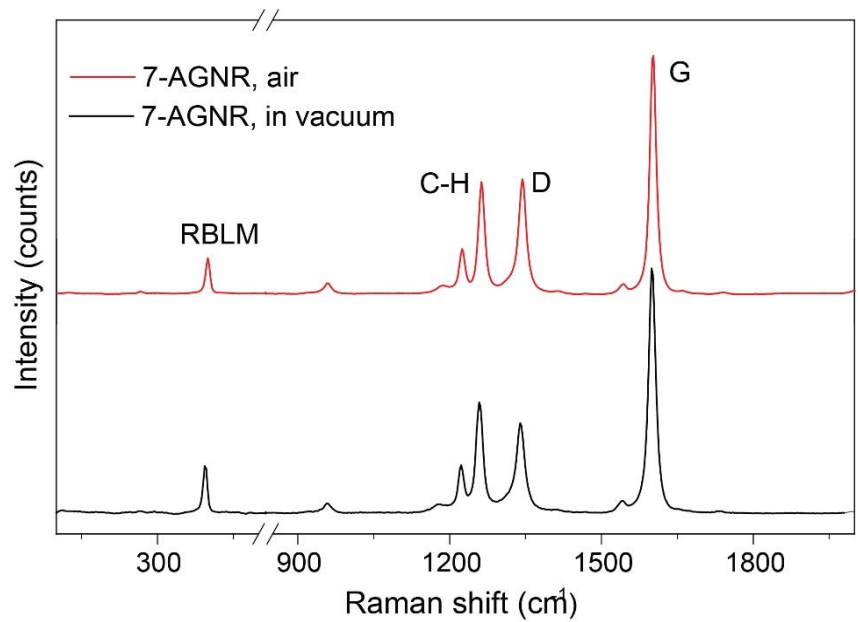

**Figure S11**. Raman spectroscopy of long 7-AGNRs in ultra-high vacuum (in black) and in air (in red) measured with 532 nm laser energy.

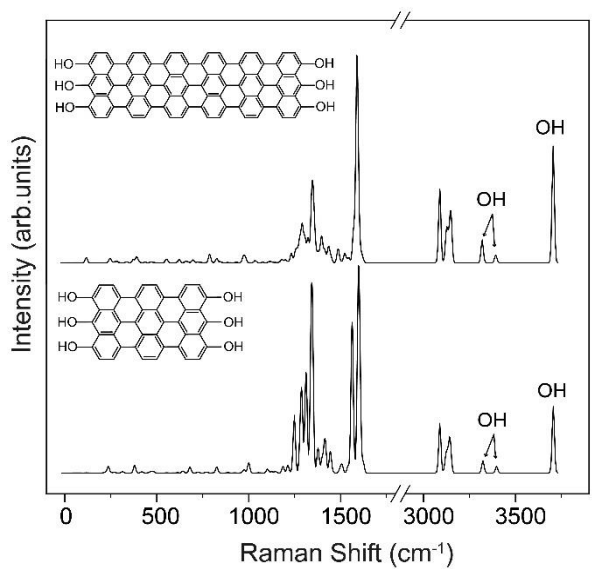

**Figure S12**. Simulated Raman spectra of HA and TA with OH-terminated zigzag edges.